# Characterization of the silicon+$^6$LiF thermal neutron detection technique

A. Pappalardo[1], M. Barbagallo[2], L. Cosentino[1], C. Marchetta[1],
A. Musumarra[1,3], C. Scirè[1], S. Scirè[1], G. Vecchio[1] and P. Finocchiaro[1]

[1] INFN Laboratori Nazionali del Sud, Catania, Italy
[2] INFN Sezione di Bari, Italy
[3] Università di Catania, Italy

**Abstract**

The worldwide need to replace $^3$He for the neutron detection has triggered R&D on new technologies and methods. A promising one is based on commercial solid state silicon detectors coupled with thin neutron converter layers containing $^6$Li. After proving the feasibility of this technique, we characterized the behavior of such a detector with different converter layer thicknesses. In this paper we also disentangle other contributions to the overall spectrum shape observed with this kind of detector, proving that its detection efficiency can be made reasonably high and that the gamma/neutron discrimination capability is comparable to the one of $^3$He tubes.

## 1    Introduction

During the last years the severe lack and the increasing cost of $^3$He have triggered a worldwide R&D program seeking new technologies and methods for the neutron detection. A viable alternative is needed to $^3$He-based neutron detectors which so far have been the most widely used systems, being almost insensitive to radiation other than thermal neutrons [1], [2]. One of the possibilities is the use of a semiconductor-based charged particle detector in combination with a neutron reactive film which converts neutrons into other detectable radiation [3]-[9]. Several principle studies have been done about this technique, and we have recently proved its feasibility by using commercial 3cm x 3cm silicon detectors coupled with thin layers of $^6$LiF, where the $^6$Li acts as converter due to its 940 b cross section to thermal neutrons [10]. The energy spectrum measured by the silicon detectors in such a configuration has a typical shape, and should allow to easily discriminate the capture reaction products from the background gamma rays.

In this paper we report on the performance of such a detector with three different thicknesses of converter layer, deposited onto two different substrates, also disentangling the contribution of the substrate itself. Other nuclear reactions occurring on the converter and on the silicon detector were also estimated, and we show that their contribution can be neglected or accounted for.

## 2    The basic test device

### 2.1    Silicon detector

The silicon detector used for our characterization and test is an MSX09-300 produced by Micron Semiconductors, a 3cm x 3cm pad, 300 μm thick, fully depleted at 30 V and therefore suitable to be coupled to neutron converters on both sides. We have already proved that this detector can be operated in such a double-converter configuration [10], but for the current tests we have chosen to use it in single-converter configuration to simplify the installation and replacement of the several substrates used. The detector, with and without a converter installed, is shown in Figure 1. The distance between the converter and the detector surface was set at about 0.2 mm, by means of four suitable plastic screws, to prevent accidental contact leading to a possible (though unlikely) fall of $^6$LiF traces onto the detector surface.

The detector output was sent to a preamplifier (Ortec 142B) and was shaped by means of a spectroscopic amplifier (Ortec 572). The amplifier output was connected to an Amptek MCA8000A multichannel analyzer in its turn connected, via USB port, to a data acquisition laptop.

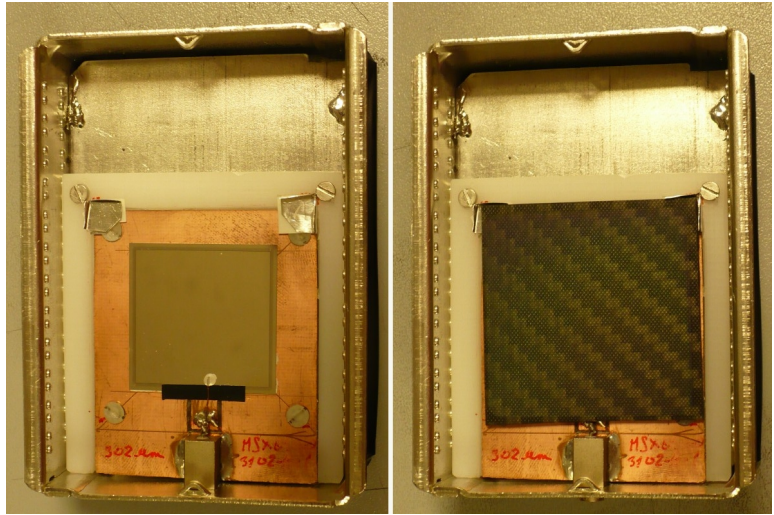

Figure 1. Left: the silicon pad detector (3cm x 3cm) used for the characterization and test. Right: the detector with a $^6$LiF neutron converter, deposited on a carbon fibre substrate, placed on top.

*2.2  The converters*

As said, the neutron converter material we chose is $^6$LiF. This material is available as a powder, enriched in $^6$Li at 95%, and we evaporated it under vacuum onto several different substrates in different thicknesses. The layer thickness was measured by means of the change in the oscillation frequency of a quartz as the $^6$LiF is evaporated onto it [11]. The reliability of the thickness measurements was checked by measuring the energy loss of alpha particles from a known source when crossing a converter layer deposited onto a suitably thin known substrate [10]. The thickness precision resulted of the order of a few percent, as well as its uniformity, checked on 5 different points. For the tests we are going to describe three different $^6$LiF layers were deposited on two different substrate types: 1.6 µm on 0.85 mm thick carbon fibre, 8 µm and 16 µm on 1.0 mm thick glass. The neutron conversion mechanism exploits the well known reaction:

$$^6Li + n \rightarrow {}^3H \ (2.73 \ MeV) + \alpha \ (2.05 \ MeV) \qquad (1)$$

which is the only possible decay channel following the neutron capture in $^6$Li, and is free of gamma rays. The energy spectrum measured by the silicon detectors in such a configuration has a typical shape, and allows to discriminate the capture reaction products (a peak from tritons and a bump from alphas) from the low-energy background due to gamma rays. In Figure 2 we reported the cross section of the reaction (1) as known from [12]. The main problem that has so far limited the use of this technique was that, in order to increase the neutron detection efficiency, one has to increase the converter thickness. This spreads the alpha and triton peaks toward the low-energy part of the spectrum and into the background, somehow compromising the gamma/neutron discrimination. We will show that a useful tradeoff can be found between thickness, discrimination and detection efficiency, which allows to obtain neutron detectors suitable for many practical applications.

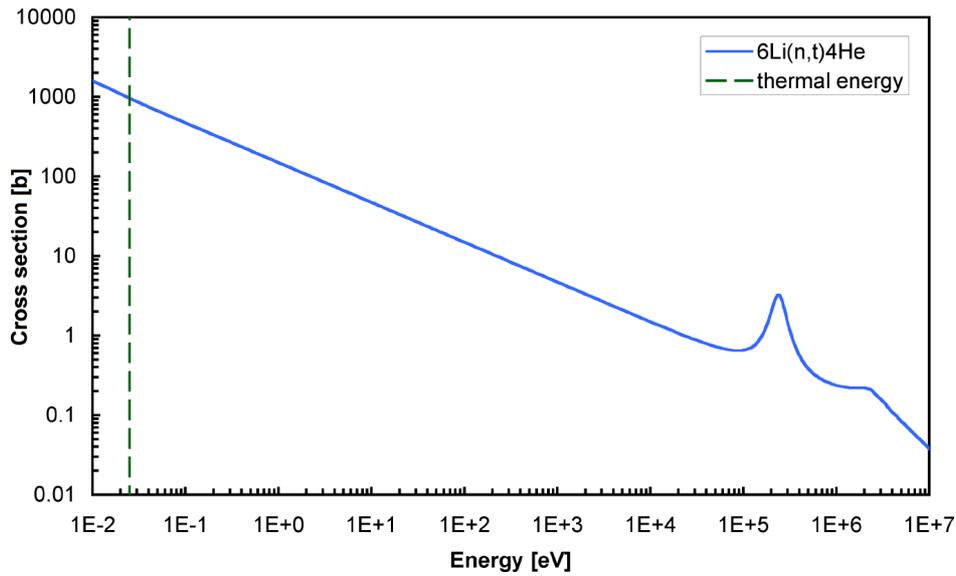

Figure 2. Cross section of the reaction $^6$Li(n,t)$^4$He.

## 2.3 Calibration of the silicon detector

Before starting any measurement we selected the electronic chain settings to be used for the whole duration of the tests, and then performed a calibration of the silicon detector by means of a mixed $^{239}$Pu+$^{241}$Am+$^{244}$Cm alpha source. Such a source produces alpha particles with three main energies (average values 5.147, 5.477 and 5.793 MeV), which we corrected to account for the energy loss due to a 2.7 mm path of the alpha particles in air before hitting the detector (for safety reasons the source case enforces this minimum distance between the radioactive material and the exit hole rim). The energy resolution thus obtained is around 200 keV FWHM, yet allowing a good calibration of the spectra into an absolute energy scale. In Figure 3 we show the calibration spectrum measured under these conditions, which was used to convert all of the following spectra to energy units.

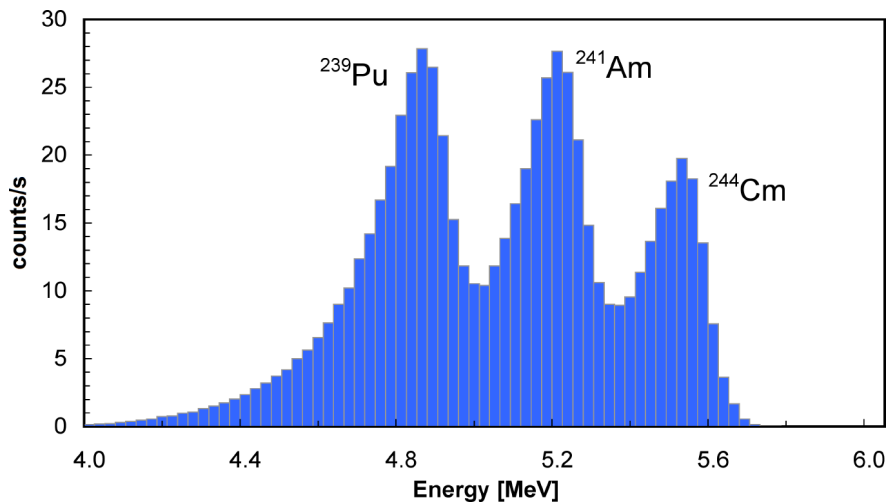

Figure 3. Energy calibration spectrum of the silicon detector. The employed non-collimated alpha source was a mix of Pu-Am-Cm. The width of the peaks (≈ 200 keV FWHM) comes from a minimum path in air of 2.7 mm due to safety reasons.

## 2.4 The test setup

In order to perform all of the needed tests the detector, alternatively with or without the appropriate neutron converter, was placed inside a box containing a polyethylene cylinder (20 cm diameter) to host the neutron source and several polyethylene slabs according to the sketch of Figure 4. The neutron source employed is an AmBe producing $1.6 \cdot 10^6$ neutrons/s, and when needed it was inserted into the hole (b) of Figure 4. The polyethylene is needed in order to soften the original neutron spectrum from the source, that extends up to 10 MeV, towards lower energy [13]. This process needs an average number of 19-20 collisions (in polyethylene, $CH_2$), corresponding to a linear path of the order of ≈20 cm for a complete thermalization. We were not aiming at setting up a precision setup for well-controlled neutron irradiation under uniform field, which would need an accurate simulation of all the materials and size, as our goal was only to enrich the spectrum with low-energy neutrons. Therefore we estimated that the 10 cm radius cylinder (8 cm effective thickness) and the use of 5 cm thick slabs as further reflector/moderator were suited to fulfil the task, as was confirmed by the data. Thicker slabs would decrease the number of neutrons due to the increasing probability of capture on Hydrogen. It has to be remarked that the AmBe source also emits a huge number of low energy gamma rays (59.5 keV from Americium) and as many high energy gamma rays as neutrons (4.4 MeV from the alpha+Be reaction).

The measurements were done in 1200, 3600 or 10000 seconds, each time in order to collect enough statistics, and this is why we normalized all of them to the data acquisition time. Therefore the data were generally plotted in counts/s. Moreover, from measurement to measurement we observed in some cases systematic shifts of the overall source-related background counting rate of the order of 10%, as slight changes in the position of detector and/or source mainly affect trajectories and solid angle of gamma rays and fast neutrons.

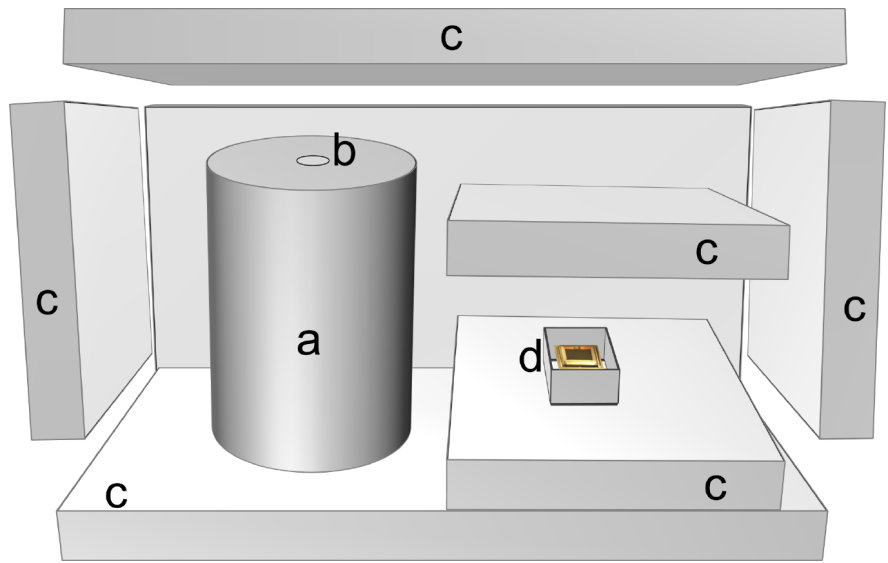

Figure 4. Sketch of the neutron moderator setup used for all the tests. (a) The polyethylene cylinder hosting the AmBe source. (b) Hole for the source insertion. (c) Polyethylene slabs (the front one is not shown). (d) The detector.

## 3   The measurements

### 3.1   Background and gamma source

The first two measurements were devoted to evaluating the behavior of the detector when exposed to the environment background and to a $^{60}$Co gamma source placed directly on top of the detector. Figure 5 shows how the background is tiny and only affecting the low-energy part of the spectrum (below 300 keV), whereas the response to the gamma source drops down by more than four orders of magnitude when going from 250 keV to the maximum $^{60}$Co gamma energy around 1.3 MeV.

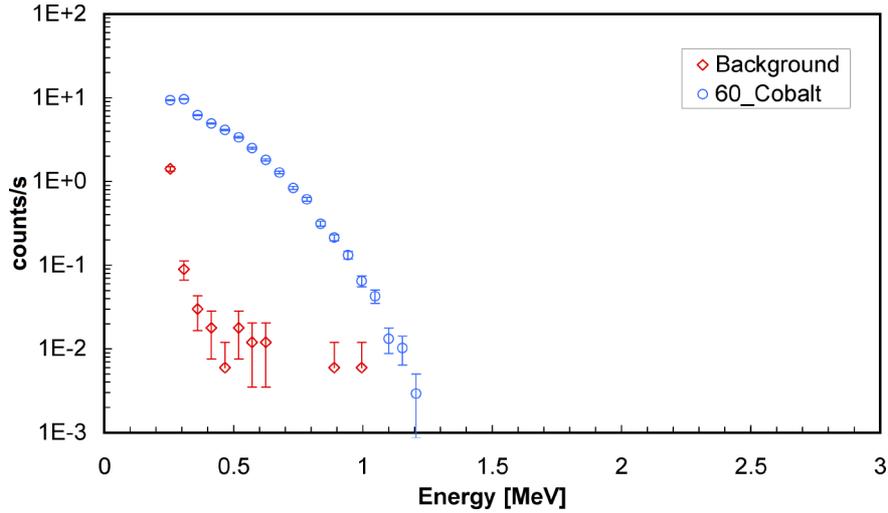

Figure 5. Energy spectrum of the silicon detector when exposed respectively to: (i) environment background; (ii) a $^{60}$Co gamma source.

### 3.2   1.6 μm $^6$LiF layer on carbon fibre

For this set of measurements we made use of the 1.6 μm thick $^6$LiF converter deposited onto the carbon fibre substrate. After placing the converter on top of the silicon detector we collected the energy spectrum with the AmBe source in place. We then performed another measurement by simply turning the converter upside down. As in this configuration neither tritons nor alphas could reach the silicon detector (the range of 2.73 MeV tritons in carbon fibre is about 50 μm), this was the perfect way to measure the contribution of the substrate. In Figure 6 (top) we show these two spectra, along with their difference which represents the contribution from the neutron converter alone. One can immediately see that this last spectrum features a clear triton peak and a wider bump due to alphas, as expected. The upper ends of these two structures are absolutely compatible with the expected values of 2.73 MeV and 2.05 MeV. The ratio between alpha and triton areas, i.e. the relative yield, is a quite reasonable 67% in line with the expectation from calculations as in ref.[5].

### 3.3   8 μm $^6$LiF layer on glass

For this set of measurements we made use of the 8 μm thick $^6$LiF converter deposited onto a glass substrate. After placing the converter on top of the silicon detector we collected the energy spectrum with the AmBe source in place. Then again, we performed the substrate background measurement by turning the converter upside down. In Figure 6 (middle) we show the two spectra, along with their difference which represents the contribution from the neutron converter alone. Again one can see the triton peak, broader than in the previous case, whereas the bump due to alphas is no longer distinguishable as it was shifted toward lower energies and into the background.

### 3.4   16 μm $^6$LiF layer on glass

With the identical procedure used in the previous cases we also tested the 16 μm thick $^6$LiF converter deposited onto a glass substrate. From the resulting plots, shown in Figure 6 (bottom), one can see that the triton peak becomes still broader and that part of the neutron efficiency is therefore not resolved and lost into the low energy background.

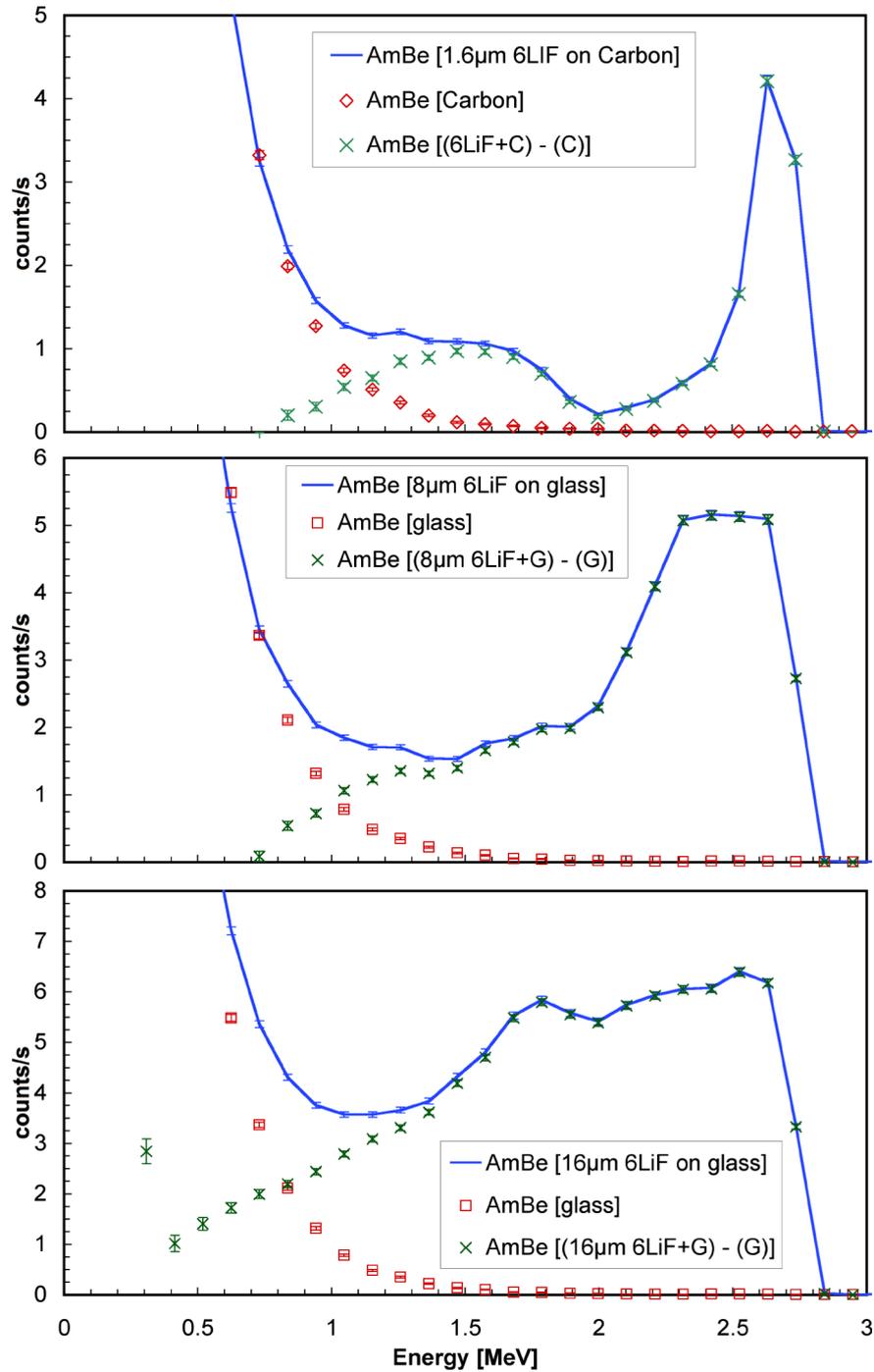

Figure 6. Top: three energy spectra with the AmBe neutron source when using respectively: (i) a 1.6 µm converter on carbon substrate; (ii) the glass substrate upside down; (iii) the difference between the previous two spectra. Middle: the same plots for an 8 µm converter on glass substrate. Bottom: the same plots for a 16 µm converter on glass substrate.

*3.5 Comparisons, results and efficiency*

In order to facilitate the comparison between the three different configurations, in Figure 7 we reported the three measured total spectra. In order to decide on the identification one has to set a threshold and decide to accept as neutrons all the detector counts above it, disregarding all the counts below. For each converter thickness one could choose a different threshold, as a tradeoff between efficiency and background. However, just to compare the efficiency gain when increasing the thickness, we set an energy window, the same for the three cases, between 1.57 MeV and 2.73 MeV. The number of neutrons/s detected under this assumption and the corresponding background counts measured with the substrate alone are reported in Table 1. By looking at these numbers one could assume that the gamma rejection power of this kind of detector is poor. As we will show further on, this is not the case as we will prove that the background counts observed with the substrate alone are not due to gamma rays.

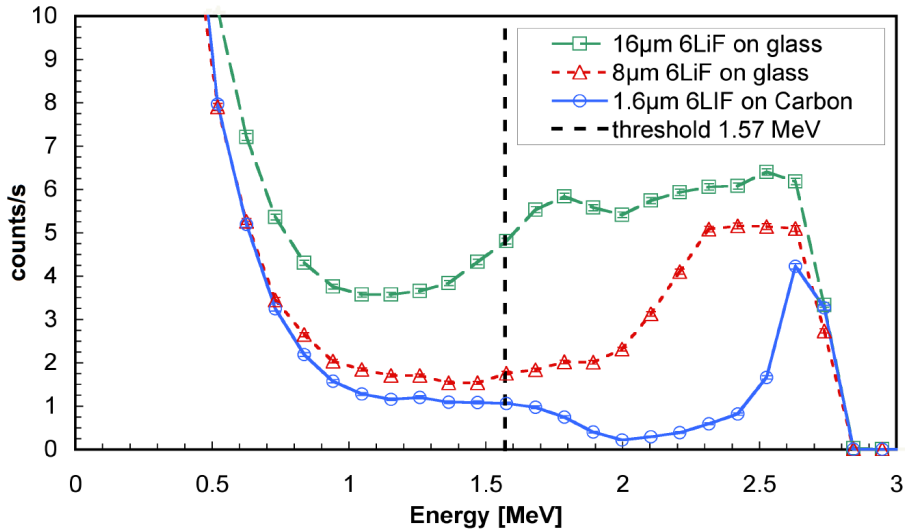

Figure 7. The three energy spectra as measured with the AmBe neutron source and the three different converter thicknesses.

| converter thickness | neutrons/s with converter | counts/s with substrate only | ratio total to background |
|---|---|---|---|
| 1.6 μm | 14.6 ± 0.11 | 0.38 ± 0.018 | ≈ 38 |
| 8 μm | 40.4 ± 0.18 | 0.35 ± 0.017 | ≈115 |
| 16 μm | 66.9 ± 0.24 | 0.35 ± 0.017 | ≈ 191 |
| bare silicon detector (no substrate) | - | 0.38 ± 0.006 | - |

Table 1. The number of neutrons/s detected with the three converter thicknesses, when setting the identification energy window between 1.57 MeV and 2.73 MeV. Also reported are the background counts measured in the same window with the substrate alone. In the fourth row the same value of counting rate for the bare silicon detector, without substrate, indicates that the substrate contribution itself is negligible. Only the statistical error was quoted.

## 4   Background beyond 3 MeV

### 4.1   *Contribution from the $^6$LiF converter*

By looking at the spectra tails at higher energy, we found out that there were unexpected counts beyond the maximum triton energy, even though about three orders of magnitude lower than tritons, up to nearly 7 MeV. The first hypothesis was that this could be a spurious contribution from the $^6$LiF layer, but it was discarded after plotting the ratio between the counting rate when the converter was facing the detector and when it was upside down. Indeed Figure 8 (left) shows the full spectrum in logarithmic scale with the AmBe neutron source, for the three converter thicknesses and when the respective substrate was turned upside down. In Figure 8 (right) we show the ratio between these two spectra, respectively for each converter thickness. A linear fit, excluding the interval where alphas and tritons mostly contribute, gives a constant ratio compatible with unity within 10%. This is a clear indication that the high energy background cannot be due to the $^6$LiF layer, as it is the same with and without the converter for the three cases.

In order to investigate the possibility that part of the $^6$LiF could have accidentally migrated from a substrate to the silicon detector surface we also built the spectrum with the AmBe source using the bare silicon, then we cleaned it and rebuilt the spectrum, and finally we tested a brand new identical detector. No significant difference was found between the three cases.

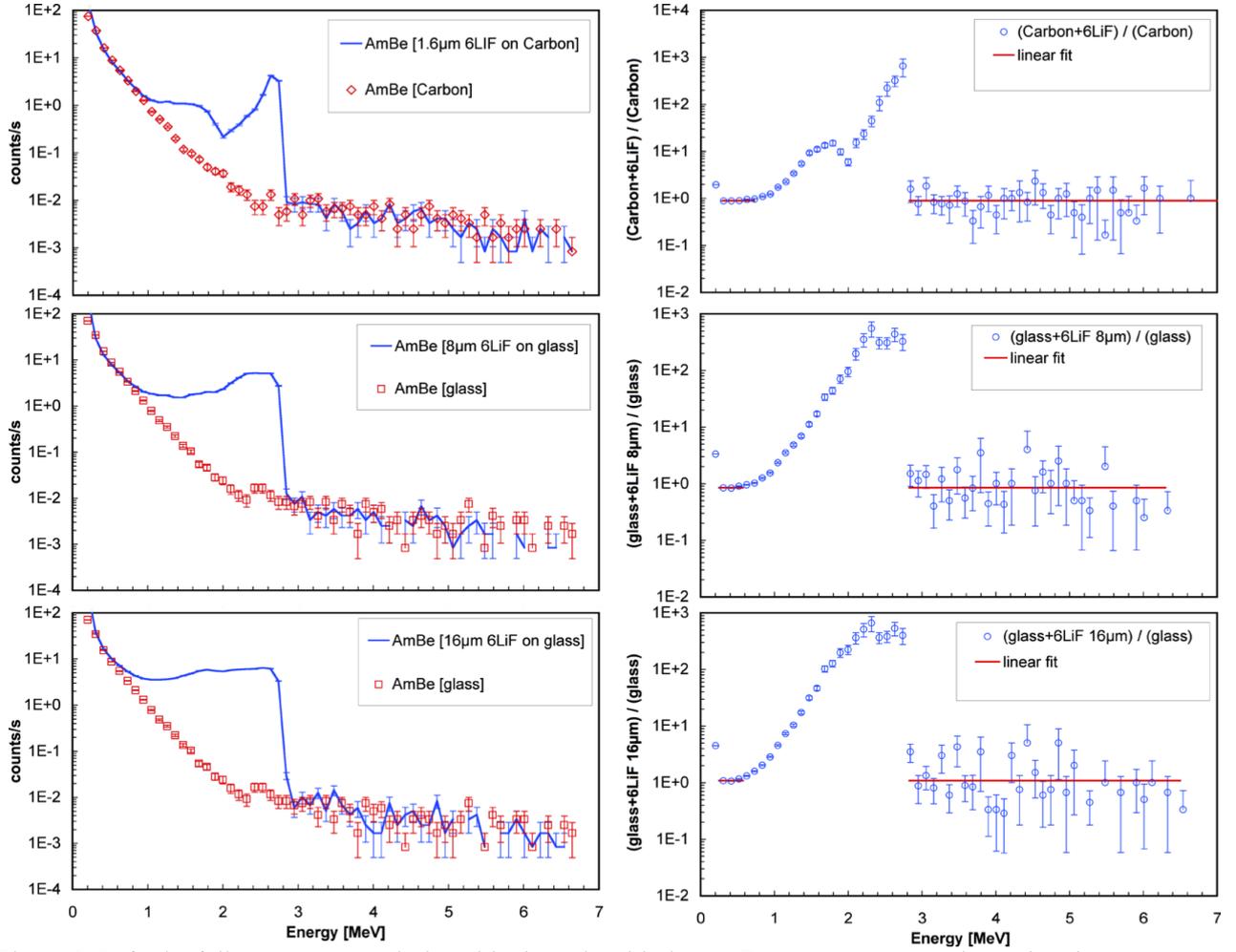

Figure 8. Left: the full energy spectra in logarithmic scale with the AmBe neutron source, when using the converter or the substrate upside down. Right: ratio between the two spectra on the left. A linear fit, excluding the interval where alphas and tritons mostly contribute, gives a constant ratio around 1. From top to bottom: converter thickness 1.6 μm (on carbon fibre), 8 μm, 16 μm (on glass).

## 4.2 Contribution from the substrate

In order to check whether the high energy background could be produced by the substrate, we plotted in Figure 9 in logarithmic scale the full spectrum taken with the AmBe neutron source when the silicon detector was respectively covered with the carbon fibre substrate, covered with the glass substrate, or bare (no substrate). The spectrum with the bare silicon was collected with a high statistics, to provide a significant content also at the highest energy values. No significant difference was found between the three cases, thus implying that the contribution from the substrate, if any, is negligible. Moreover, the counting rate in the same energy range as in Table 1 was $0.38 \pm 0.006$ counts/s, identical to what observed with the substrate in place.

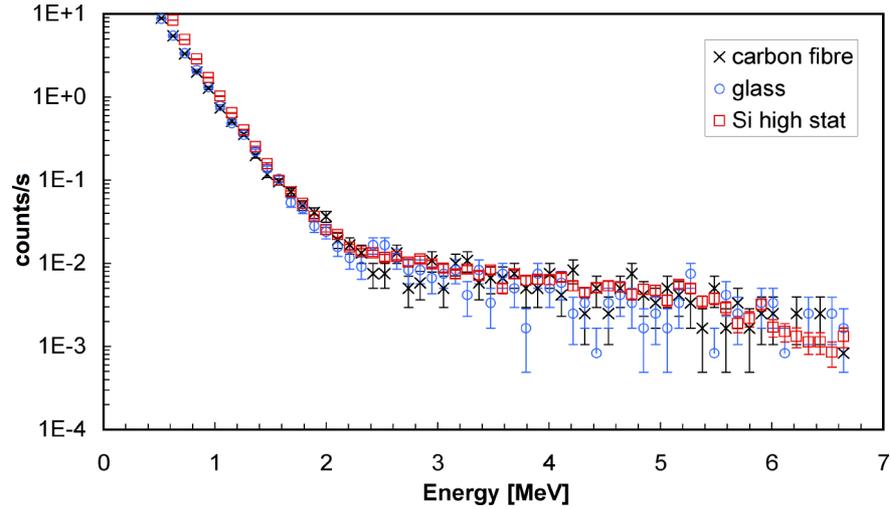

Figure 9. The full energy spectra in logarithmic scale with the AmBe neutron source, when using respectively: (i) a carbon fibre substrate; (ii) a glass substrate; (iii) no substrate, i.e. bare silicon detector. No significant difference was observed, implying that the substrate contribution is negligible.

*4.3  Contribution from other mechanisms*

In order to understand the origin of the higher energy part of the measured spectrum, we investigated other possible mechanisms capable of depositing enough energy into the silicon detector. Cosmic muons are excluded, as their energy loss in the detector is quite small and indeed in Figure 5 there are no counts above 1 MeV. Another possibility could be high energy gamma rays, either a chance coincidence of two 4.4 MeV gammas directly coming from the AmBe source (very unlikely), or others from (n,gamma) reactions on the materials surrounding the detector. This second case is restricted to Silicon, as it is the only material present having reasonable cross section at thermal neutron energy ($\approx 0.1$ b), and that produces a cascade with two gamma rays of respectively 4.9 MeV and 3.5 MeV.

We simulated the response of the detector when these two gamma rays are produced simultaneously in a random position inside it. Assuming a realistic number of incoming thermal neutrons, the simulation result is about three to four orders of magnitude lower than experimentally measured, as can be seen in Figure 10. Moreover, the simulation basically excludes the possibility of gamma events with large energy deposit up to 6-7 MeV.

In order to disentangle other possible mechanisms capable to account for those events, we looked up the cross sections for the materials somehow present in the vicinity of the detector into the ENDF/B-VII.1 data base. We assumed that such a high energy could come from protons in (n,p) reactions produced by a tail of high-energy neutrons from the AmBe source. As the detector was enclosed in a small steel box 0.6 mm thick, protons from outside it were ruled out as they would be stopped in any case. Simple considerations about geometry, thickness and cross sections of the materials basically ruled out Silicon, Iron and Copper, with a possible tiny contribution from (n,p) elastic scattering on Hydrogen from the PCB board.

Therefore we assembled a sandwich of two identical detectors without neutron converter, placed it inside the moderator and collected the 2-dimensional energy scatter plot shown in Figure 11 by triggering on at least one detector fired. The region (a) corresponds to events where a neutron was scattered elastically off silicon nuclei on the detectors, the measured energies are due to the recoiling silicon nuclei. The regions (b) and (c) represent events where only one detector was involved, with the neutron reacting with a silicon nucleus and most of the available kinetic energy deposited into the detector. The regions (d) and (e) represent events where a neutron is first elastically scattered in one detector and then undergoes the same interaction as in (b,c) in the other detector. We also observe that in (e) the energy of the recoiling nucleus is larger than in (d), as expected from kinematic considerations (backscattered neutron and maximum energy transfer). The region (f), with the typical shape produced by a silicon telescope and a tiny number of counts, is due to (n,p) scattering from Hydrogen in the PCB board supporting the detector, as said before.

Indeed, the involved cross sections are realistically in agreement with our interpretation of Figure 11. A precise simulation could likely corroborate this interpretation, but it is not strictly needed as the only important point was to exclude that the high energy contribution to the measured spectra could come from gamma rays. This was proved, and we can conclude that the silicon detector itself also shows some low sensitivity to high energy neutrons.

The shape of the "background" silicon spectrum of Figure 9 and Figure 10 is therefore due to three different components: (i) low (deposited) energy gamma rays; (ii) (multiple) elastic scattering of neutrons on silicon, up to ≈ 1.5 MeV; (iii) neutron interaction on silicon up to the maximum possible neutron energy. None of these mechanisms can perturb the information conveyed by the detector as neutron counter, because the additional counts are still due to neutrons, while still keeping the optimum neutron/gamma discrimination power.

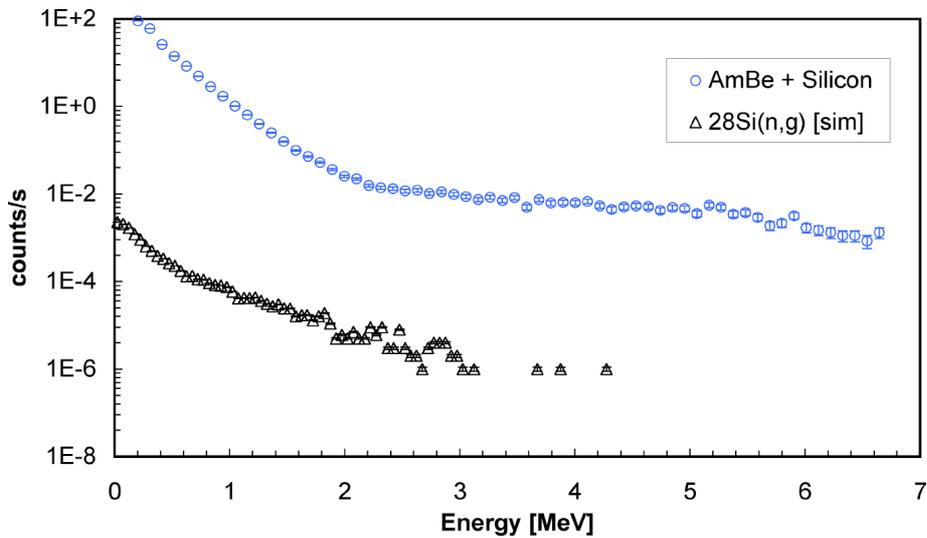

Figure 10. Triangles: simulation of the spectrum in the silicon detector by 4.9 and 3.5 MeV gamma rays simultaneosuly produced inside it. Circles: the experimental spectrum measured with the AmBe source and the bare silicon detector.

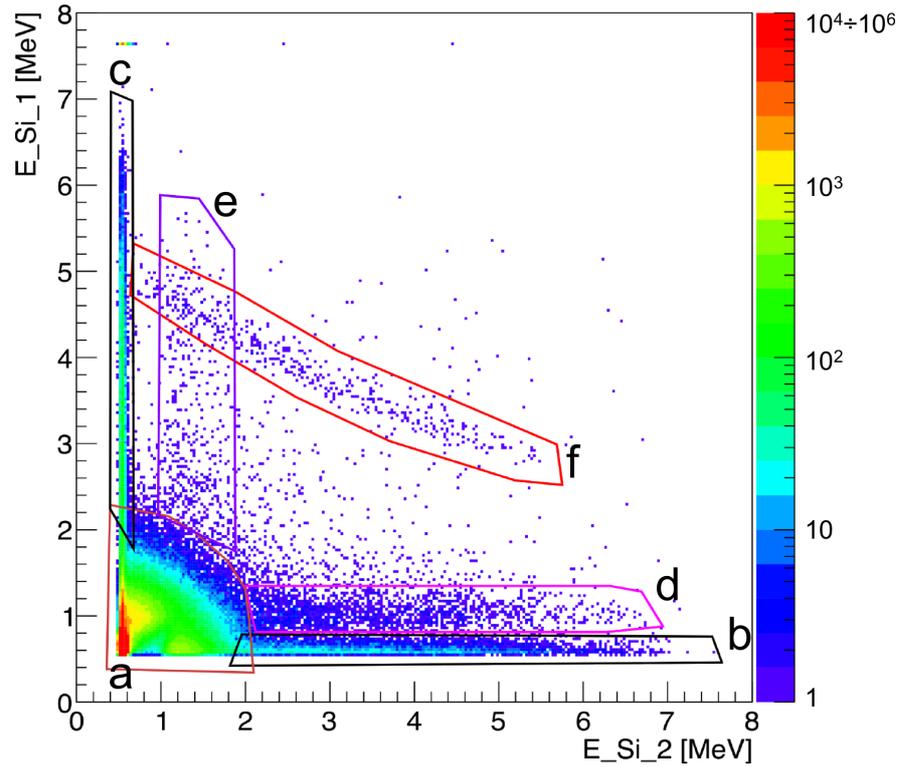
Figure 11. Energy versus energy scatter plot for the 2-detector sandwich. See the text for details.

## 5   Discussion

The data presented in this paper clearly show that a layer of $^6$LiF can fruitfully be used as neutron converter, to be placed on top of a solid state silicon detector for detection and identification of thermal neutrons. A background contribution was found to be present, but it has to be ascribed to the interaction of fast neutrons coming from the source with the silicon detector itself. This means that the detector is also slightly sensitive to fast neutrons, which are still neutrons even though of much higher energy, and by means of different interaction mechanisms. However, a better thermalization can reduce, if not suppress, these counts. As for the sensitivity to gamma rays, we have seen that the efficiency around 1 MeV to a $^{60}$Co source is $< 10^{-6}$. The simulation with high energy gamma rays, 4.9 MeV, shows that the efficiency in the region E > 1 MeV is of the order of $5\times10^{-6}$. Therefore from the gamma rejection point of view the detector is to be considered equivalent to a $^3$He tube.

In order to evaluate the neutron detection efficiency one can make use of the 1.6 μm thick converter, and select events in the triton peak only in Figure 6 (top), i.e. in an energy window from 2 MeV to 2.73 MeV, because in such a case this peak is well resolved. The amount of $^6$Li in the converter is well known, as well as its thermal neutron cross section, and therefore under these well defined conditions one obtains a value of 0.42% for the expected neutron detection efficiency (including the angular inefficiency do to the possible emission of the triton at large angles from deep inside the converter [5]). Using this value of the efficiency we could estimate the rate of thermal neutrons impinging on the detector, that was about 2700/s.

In light of all these evidences, one can immediately evaluate the performance of the detector when two 16 μm thick coverter layers are used, one on each detector face, and of a stack of two such detectors. In Figure 12 we show the detection efficiency of these two configurations as a function of the lower energy threshold, while keeping the upper threshold at 2.73 MeV (for the stack the absorption in the first detector was taken into account). We also evaluated the purity of the neutron information under our experimental conditions, i.e. the fraction of counts effectively produced by the converter, thus not including the counts due to neutron reactions in the silicon which however one could consider detected neutrons as well. If setting the lower threshold at 1.5 MeV the efficiency of the two-converter detector is 5.2%, whereas the stack reaches up to 10%, with a purity of 99.3%.

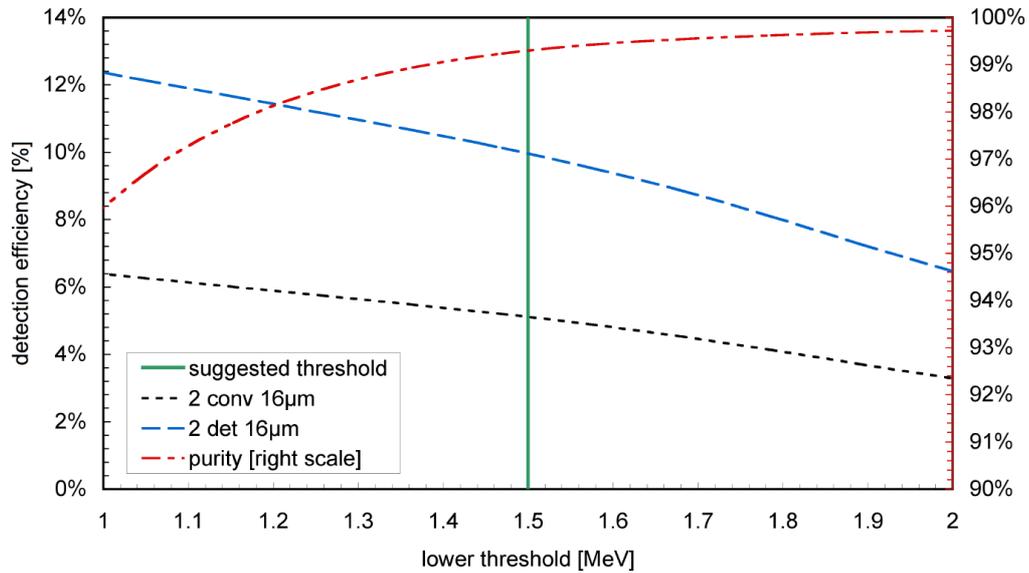

Figure 12. Efficiency as a function of the lower energy threshold used, the upper threshold is 2.73 MeV. Two configurations were evaluated: (i) a silicon detector with two 16 μm thick converters (one per face); (ii) a stack of two such detectors. The purity of the slow neutron signal under our experimental conditions was also reported, to be read on the righthand side scale.

## 6 Conclusion

In the framework of the worldwide R&D programs seeking new technologies and methods for the neutron detection we have highlighted a viable alternative basically insensitive to gamma radiation. The semiconductor-based charged particle detector, coupled with a thin $^6$LiF neutron converter, has shown a reasonable detection efficiency up to 5.2%. The measurements on such a detector with three different thicknesses of converter layer allowed us to carefully study its performance and to disentangle the contribution of the substrate. Other nuclear reactions possibly occurring on the converter and on the silicon detector were also taken into account, and we showed that their contribution can be neglected or accounted for.

The 16 μm $^6$LiF layer thickness could still be slightly increased, thus gaining in efficiency. An efficiency leap is achieved by stacking two or more of these detectors, while keeping the gamma efficiency around $10^{-6}$, that basically corresponds to the same n/gamma discrimination power obtained with $^3$He tubes. Indeed 10% efficiency can be obtained with a stack of two detectors each equipped with two 16 μm thick converters.

The flat geometrical structure of the detector allows an easy evaluation of its efficiency, once the $^6$LiF enrichment and thickness are known, as the $^6$Li(n,alpha)t is a well known standard cross section. Moreover, the flatness also facilitates installation and use, while keeping a constant efficiency throughout its entire area. Worth to be mentioned is also the low voltage operation, which simplifies the detector use quite a lot, as well as the very simple mechanical structure and the absence of gas for its operation. Several applications of the proposed detection technique are already operational, like for instance at the n-TOF facility [14],[15],[16], and others will likely follow quite soon.

Last but not least, very recently we have also checked the performance of such a detector, as compared to $^3$He tubes, in a neutron diffractometer installed on a time of flight beam line at a spallation source. The results were outstanding and will be shown in a forthcoming paper [17].

# 7 Acknowledgments

We are grateful to our friend and colleague Marco Ripani, responsible of the INFN-Energy project, for the constant support to our developments in neutron detection techniques. We are also strongly indebted with Eugenio Costa for his valuable contribution in the production of the $^6$LiF converters.